# High-Efficiency, Extreme-Numerical-Aperture Metasurfaces Based on Partial Control of the Phase of Light

*Claudio U. Hail, Dimos Poulikakos,\* Hadi Eghlidi\**

Laboratory of Thermodynamics in Emerging Technologies, ETH Zürich, Sonneggstrasse 3, CH-8092 Zürich, Switzerland. E-mail: dpoulikakos@ethz.ch, eghlidim@ethz.ch

**Abstract**
High-quality flat optical elements require efficient light deflection to large angles and over a wide wavelength spectrum. Although phase gradient metasurfaces achieve this by continuously adding phase shifts in the range of 0 to $2\pi$ to the electric field with subwavelength-sized scatterers, their performance is limited by the spatial resolution of phase modulation at the interface. Here, we introduce a new class of metasurfaces based on a general formulation, where the phase shifts cover less than the full 0-$2\pi$ range, offering significant advantages. More specifically, this approach allows the realization of metasurfaces with more compact and less mutually-interacting scatterers, thus more precise phase modulation, and advances the performance limits of metasurfaces to domains significantly beyond those of the full coverage phase gradient approach. Applying this concept to both plasmonic and dielectric surfaces, we demonstrate large phase gradients resulting in high-numerical-aperture immersion metalenses (NA=1.4) with near diffraction-limited resolution (~0.32$\lambda$) at visible wavelengths. Our concept enables added functionalities such as a broadband performance and wavelength de-multiplexing on a single layer, surpassing the theoretical cross-polarization transmission efficiency limit for single-layer plasmonic metasurfaces, and yields 67% efficiency for dielectric metasurfaces. This work paves the way toward realizing high-resolution flat optical elements and efficient plasmonic metadevices.

**Introduction**
Deliberate control over the propagation of light is of importance to many fields including imaging, sensing, communication and energy.[1–3] Planar metamaterials enable a large degree of control by changing the amplitude, phase and polarization of light abruptly on a subwavelength scale at an interface. With these so-called metasurfaces, a multitude of ultrathin flat optical elements such as beam deflectors,[4,5] flat lenses[6–8] and polarization beam splitters[9,10] have been demonstrated. Deflecting light to large angles and efficiently manipulating its wavefront are two key requirements for these metasurface-based, flat optical elements. This is achieved with anomalous refraction of light by phase gradient metasurfaces, which continuously add phase shifts ranging from 0 to $2\pi$ to the electric field at the interface via nanoscale scatterers.[4] However, refracting light to large angles requires the ability to introduce large phase differences over a short distance. In single-layer metasurfaces these phase gradients are introduced by varying the resonance phase or the Pancharatnam-Berry (PB) phase. While the former method uses V-shaped plasmonic antenna modes[4,5,11] or Mie type scattering of high-index dielectric structures,[12,13] the latter employs simple (rod-shaped) plasmonic antennas[14,15] or propagation through dielectric structures[8,16] (see Figure 1a and b). Recently, metasurfaces based on low-loss dielectric structures have demonstrated high anomalous transmission efficiencies.[12,16,17] However, in plasmonic metasurfaces the required polarization conversion intrinsically limits the corresponding transmission efficiency of these surfaces to below 25%,[18] with the largest measured efficiency of 15% for focusing light.[19] Moreover, in both plasmonic and dielectric surfaces, it is challenging to efficiently introduce large phase gradients for large angle refraction, since the scatterer size and near-field interactions restrict the distance between phase-shifting elements.[20] Diffractive flat optical elements, while functionally similar, have their own working principle and limitations for large angle deflection or focusing (multiple real and virtual focal points, limited efficiency).[21–23]

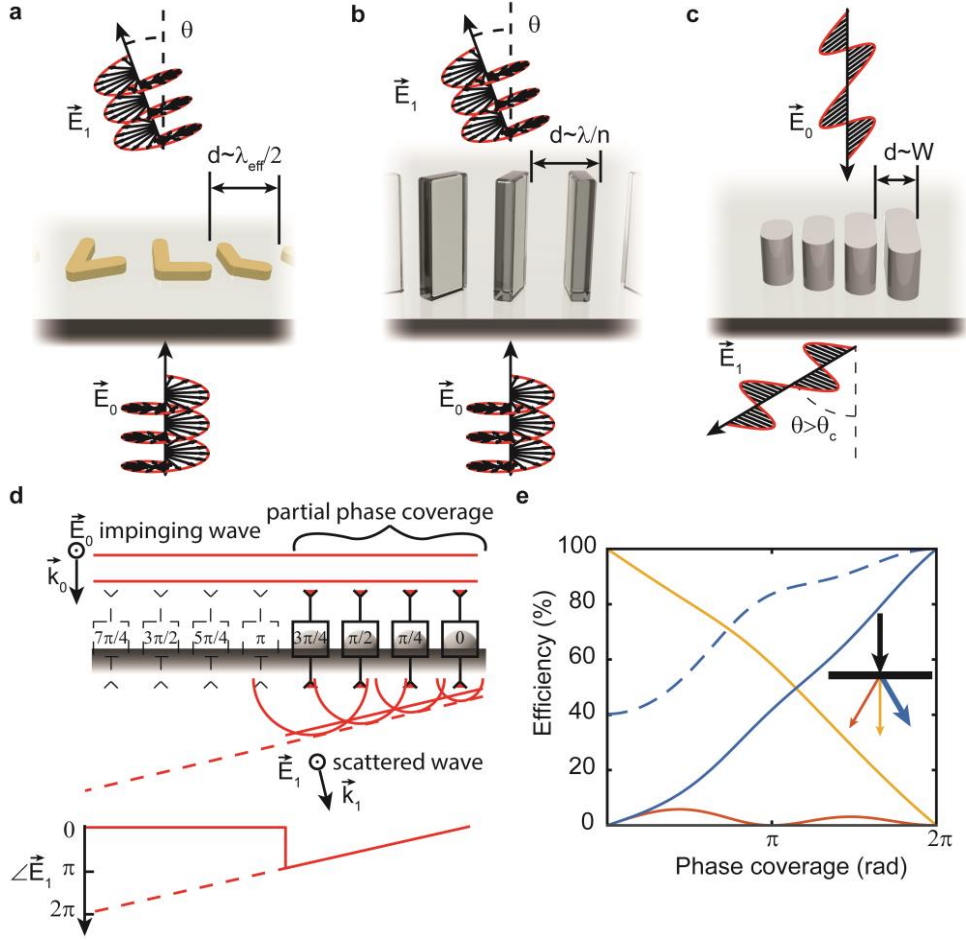

*Figure 1.* Phase gradient metasurfaces with full and partial phase control. (a) A plasmonic phase gradient metasurface for anomalous refraction illuminated with circularly polarized light.[4] The separation d between the elements is approximately given by half the effective wavelength $\lambda_{eff}$[24] and transmission efficiencies to the anomalous order are limited due to polarization conversion.[18] (b) A dielectric phase gradient metasurface with similar illumination as in (a).[8] The separation between elements can be approximated by $d \approx \lambda/n$.[25] (c) A gradient metasurface with partial phase control. The separation between the elements is comparable to the width W of the scatterer, which can be as small as a few tens of nanometers. Deflection to angles $\theta$ larger than the critical angle $\theta_c$ is possible. $E_0$ and $E_1$ represent the impinging electric field and the scattered electric field to the anomalous order, respectively, and $k_0$ and $k_1$ represent the respective wavevectors. (d) Schematic representation of the impinging and the scattered wavefronts for one period of a metasurface with partial (solid line) and full phase coverage gradient metasurface (dashed and solid line combined). The squares represent scatterers at the interface, which add phase to the impinging wavefront. $\angle E_1$ is the phase added to the wavefront at the interface. (e) Influence of partial phase control on the normal (yellow curve) and anomalous (blue curve) transmission efficiencies and the suppression of the opposite order (red curve) based on our analytical model. The dashed line represents the efficiency in the limiting case where the entire incident light interacts with the phase-shifting scatterers (for a description of the analytical model see Supporting Information).

In this work, we show how a high degree of wavefront manipulation can be achieved with only a partial control of the phase (i.e. less than 0-2π, see Figure 1c and d). With this relaxed phase requirement, the choice of scatterers is less restrictive. As a result, more compact scatterers, for example simple nanorods (Figure 1c), with reduced near-field interactions can be used to introduce large phase gradients by placing them at smaller distances from each other. This allows, in principle, accessing phase gradients for anomalous refraction at extreme angles, up to 90° in air or in dielectrics, otherwise not feasible with full phase coverage surfaces.[26] Furthermore, these metasurfaces do not rely on polarization conversion and can generate scattered light with the same polarization as the impinging light (i.e. the co-polarized beam). In contrast to single-layer plasmonic phase gradient surfaces, there is thus no limitation in efficiency due to a polarization conversion.[18] One period of a phase gradient metasurface with partial phase control is shown schematically in Figure 1d. Only the elements which provide smaller phase shifts (here less than π) are considered and the elements for larger phase shifts, which are required for conventional phase gradient metasurfaces, are omitted (elements with dashed outline in Figure 1d). The effect of the phase coverage of the scatterers on anomalous refraction can be analyzed by approximating the metasurface as a phase grating that periodically adds a linearly varying phase to the electric field (see Figure 1d). With this assumption, we calculate the transmission efficiency of the different orders as a function of the phase coverage of the surface (solid lines in Figure 1e). Here, we assume that only the part of light which impinges on the surface covered with phase shifters (elements with solid outline in Figure 1d) interacts with the metasurface (for details see Supporting Information). A single resonance metasurface with a phase coverage of π could attain anomalous refraction at efficiencies exceeding 40% and up to 84% if the entire light interacts with the phase-shifting scatterers (dashed line in Figure 1e). Moreover, similar to phase gradient metasurfaces with full phase coverage (0-2π), the contribution to the opposite diffraction direction (the red arrow in Figure 1e) is entirely suppressed. Our concept can be viewed as a hybrid approach between phase gradient metasurfaces and conventional Fresnel zone plates. While in one part of the surface the phase is controlled on a subwavelength scale (similar to phase gradient metasurfaces), in the other part the phase is not modulated (similar to Fresnel zone plates, see Fig. 1d).

We demonstrate our concept for wavefront manipulation based on partial phase control with a metasurface composed of ultrathin (45 nm thick) plasmonic linear phased arrays shown schematically in Figure 2a. Phase is added to the impinging wavefront using the localized surface plasmon resonance (LSPR) scattering of silver nanorods on a glass substrate. The wave impinging on the surface is linearly polarized along the nanorod long axis and is scattered into the substrate with the same polarization. Varying the length of the nanorods sets the phase delay between the impinging and scattered wave due to the shape and size dependence of the LSPR, as shown in Figure 2b. Additionally, the width of the nanorods can be varied to allow for more degrees of freedom in adjusting the phase and amplitude of the scattered light. The single mode LSPR of the nanorods can yield only a partial phase coverage of up to π. Arranging two or more nanorods in a subwavelength-spaced linear array introduces a piecewise linear phase profile along the interface. The distance between two elements in the array with a phase difference of $\Delta\varphi$ is given by

$$d = \Delta\varphi \frac{\lambda}{2\pi n \sin\theta}, \tag{1}$$

and ensures a phase gradient resulting in anomalous refraction at the angle θ with respect to the direction normal to the substrate with refractive index *n* and at the design wavelength λ.

Reproducing this array periodically in two directions with periodicities $P_x$ and $P_y$, creates a continuous metasurface, as shown in Figure 2a. The periodicity $P_x$ is obtained from Equation (1) by setting $\Delta\varphi = 2\pi$. As apparent from Equation (1), anomalous refraction at large angles requires a very short distance between the elements. For example, at $\lambda = 660$ nm and with a phase difference of $\pi/2$ between two elements, attaining refraction to $\theta = 60°$ in glass requires $d = 125$ nm. Achieving this distance consistently between neighboring scatterers of different kinds (e.g. dielectric nanofins or V-type nanoantennas) is very difficult due to the dimensions of the scatterers and the near-field coupling between them. However, in our linear phased array these distances are feasible due to the weak interaction between the elements and their small widths as compared to the wavelength.

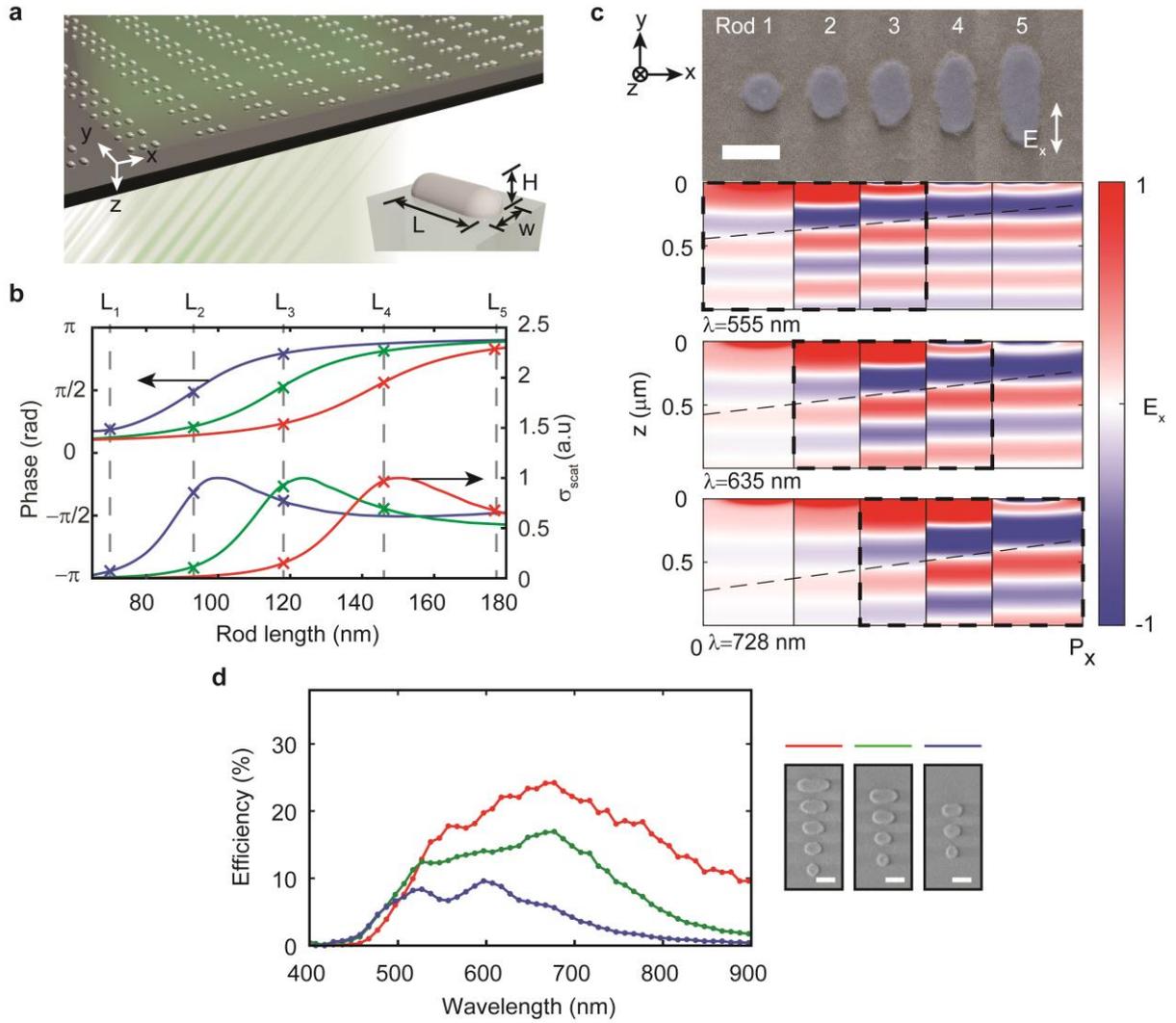

*Figure 2. Metasurface design and characterization. (a) Illustration of a three-element array metasurface for anomalous refraction of light to one direction. The inset shows a close-up illustration of a nanorod on the metasurface with length L, width W and height H. (b) Calculated phase shift and scattering cross section $\sigma_{scat}$ of a silver nanorod on a glass substrate with varying rod length at the wavelengths of 555 nm (blue curves), 635 nm (green curves) and 728 nm (red curves), for W = 70 nm. The cross symbols mark the rod lengths $L_{1-5}$ and the respective scattering intensities of the nanorods of the array in (c) and (d). (c) Calculated electric field scattered into the substrate by a five-element nanorod array at*

*the resonance wavelengths of rods 2, 3 and 4. The dashed lines show the anomalously refracted wavefront and the bold dashed lines mark the strongest interacting rod triplet at the respective wavelength. Upper panel: False colored SEM of the corresponding five-element array with the electric field polarization indicated. (d) Measured transmission to the anomalously refracted beam for three-, four- and five-element array surfaces. Nanorod sizes are $L_{1-5}$ = 70, 93, 118, 146, 177 nm, W = 70 nm and H = 45 nm. The periodicities are $P_x$ = 660 nm and $P_y$ = 300 nm. Scale bars, 100 nm.*

We fabricated metasurfaces by electron beam lithography on an evaporated silver film and used physical argon ion etching for pattern transfer. A multiple of nanorods can be used to form a metasurface with partial phase control, with a minimum requirement of two for creating a phase gradient. The nanorod dimensions are selected based on the phase delay and scattering amplitude, as shown for a varying rod length in Figure 2b for three different wavelengths. We obtain a narrow-band design by setting the middle rod to be at resonance and the two adjacent rods to provide fixed lagging or leading phase differences. In the example shown, rods 1, 2, 3 at a design wavelength of 555 nm provide a phase difference of $\Delta\varphi = 0.3\pi$ between the neighboring rods. To extend the bandwidths of this surface, more rods (4 and 5) can be added, such that rods 2, 3, 4 and rods 3, 4, 5 form the same phase gradient as above at wavelengths of 635 nm and 728 nm. Figure 2c shows a scanning electron micrograph (SEM) of a unit cell of a fabricated metasurface with nanorods 1-5, and the corresponding electric field scattered into the substrate by each nanorod calculated with finite-difference time-domain (FDTD) simulations. The distance between elements was chosen slightly larger than calculated from Equation (1) due to near-field coupling effects observed in numerical simulations. As the simulation results in Figure 2c show, at each design wavelength only the three corresponding rods scatter light intensely and bend it to the anomalous refraction direction. Thus, the four- and five-element arrays, exhibit a multi-resonant behavior and, therefore, a broadband performance.

**Results and Discussion**

We measured the transmission efficiency of the anomalous refraction of light normally incident on our metasurfaces by Fourier imaging of the transmitted light on a home-built microscope (for details see Experimental Section and Figure S1, Supporting Information). Figure 2d shows the anomalous refraction efficiency of metasurfaces consisting of three-, four- and five-element arrays. Increasing the number of elements in an array expands the full width at half maximum (FWHM) bandwidth from 200 nm to 310 nm. With a five-element array, anomalous refraction almost over the entire visible spectrum is achieved. The measured angles of anomalous refraction extend up to 63.4° in the glass substrate; larger angles cannot be fully captured by our oil immersion objective. By further reducing the periodicity along the array ($P_x$ in Figure 2c), larger angles can be attained even in the visible spectrum.

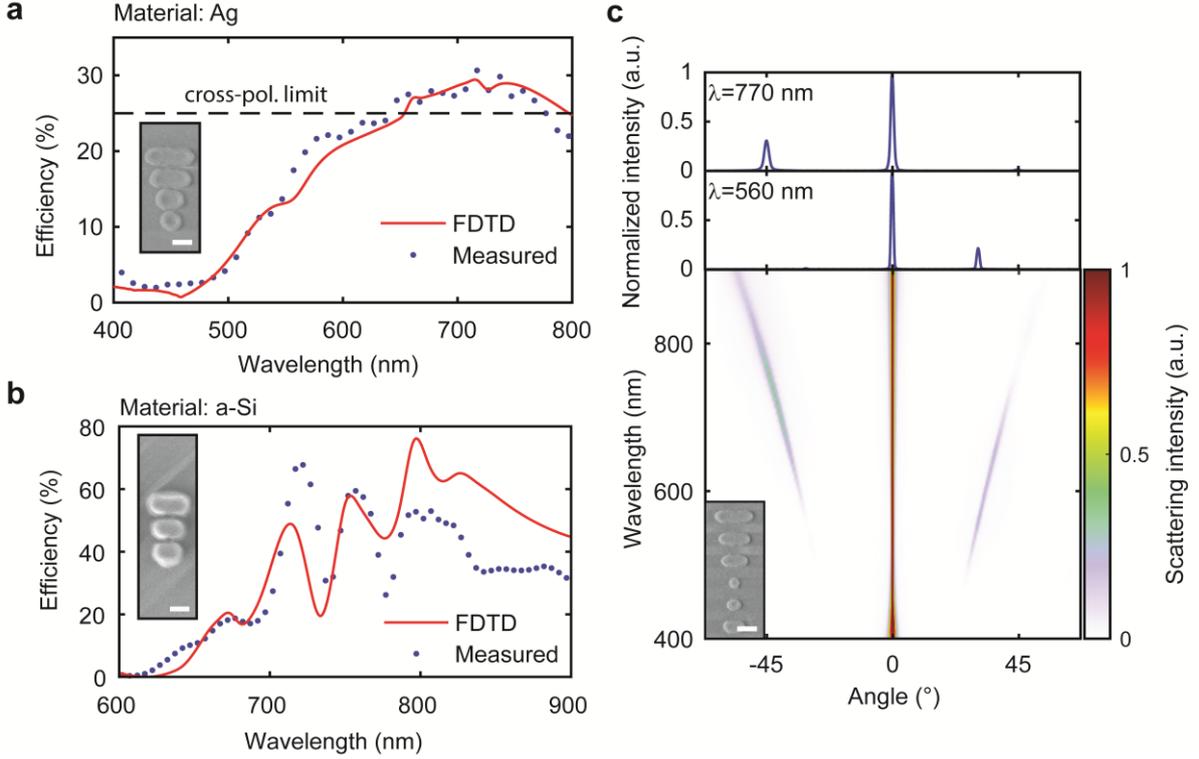

*Figure 3. Efficient broadband anomalous refraction and wavelength de-multiplexing. (a) Measured transmission efficiency of anomalous refraction of the improved plasmonic design and comparison to FDTD simulation. The dashed line shows the theoretical limit for transmission to the cross-polarized beam using plasmonic single-layer metasurfaces.[18] The periodicities of this design are $P_x = 660$ nm and $P_y = 300$ nm. (b) Measured transmission efficiency of a dielectric metasurface with partial phase control and comparison to FDTD simulation. The surface consists of an array of 350 nm tall amorphous silicon nanopillars and has a periodicity of $P_x = 950$ nm and $P_y = 300$ nm. The periodic modulation in transmission is due to interference effects within the 850 nm thick polymer layer, which covers the structures. (c) Measured transmitted scattering intensity of a plasmonic wavelength de-multiplexing metasurface as a function of wavelength and deflection angle. The periodicities of the metasurface are $P_x = 700$ nm and $P_y = 300$ nm. The top panels show the scattering intensity over different angles for wavelengths of 560 nm and 770 nm. The insets show scanning electron micrographs (SEM) of a unit cell of the measured surfaces. Scale bars, 100 nm.*

By simultaneously varying the width and length of the individual nanorods, we realized a surface with increased anomalous refraction efficiency as shown in Figure 3a. This increased efficiency results from the added degree of freedom of the width of the nanorod, which allows optimizing the scattering amplitudes of the nanorods in addition to their phase shifts. We employed full-wave simulations to account for the effects of the phase coverage, scattering cross section, scattering pattern and near-field interaction. On the resulting design, we measured a maximum transmission efficiency of 30% at a wavelength of 720 nm (see Figure 3a) in excellent agreement with predictions from simulations. This measured maximum efficiency exceeds previously reported values for plasmonic single-layer metasurfaces[27] by a factor of 2 and, thanks to scattering light to the co-polarized beam, this value is even higher than the 25% theoretical anomalous transmission efficiency limit for conventional single-layer plasmonic phase gradient metasurfaces[18] (dashed horizontal line in Figure 3a). Additionally, our simulations suggest that absorption in the metal is almost negligible (5%), and therefore, the efficiency of our surfaces is practically only limited by the scattering cross section of the scatterers. By employing dielectric scatterers[28] the efficiency can be further increased. The measured efficiency of a three-element dielectric metasurface based on our concept is shown in Figure 3b, where a maximum efficiency of 67% is obtained with a phase coverage of only $0.7\pi$ (for measurements of metasurfaces with shorter periodicities see Figure S2, Supporting Information). This shows that with efficient light interaction and the negligible losses in dielectrics, efficiencies close to the theoretical limit (dashed blue curve in Figure 1e) can be achieved.

In addition to demonstrating efficient wavefront manipulation, metasurfaces with partial phase control naturally bring about the possibility of independently realizing added functionalities on one surface as a result of the implicit partial coverage of the surface area. One example is shown in Figure 3c, where we combine two plasmonic arrays with opposing phase profiles at different wavelengths. Here, light of different wavelengths is anomalously refracted to opposite directions, resulting in wavelength de-multiplexing. Other possible realizations are phase gradient surfaces with different responses for different polarizations. Like other approaches to multifunctionality, metasurfaces with partial phase control could be designed to achieve spin-dependent[29] or multi-wavelength functionalities.[30]

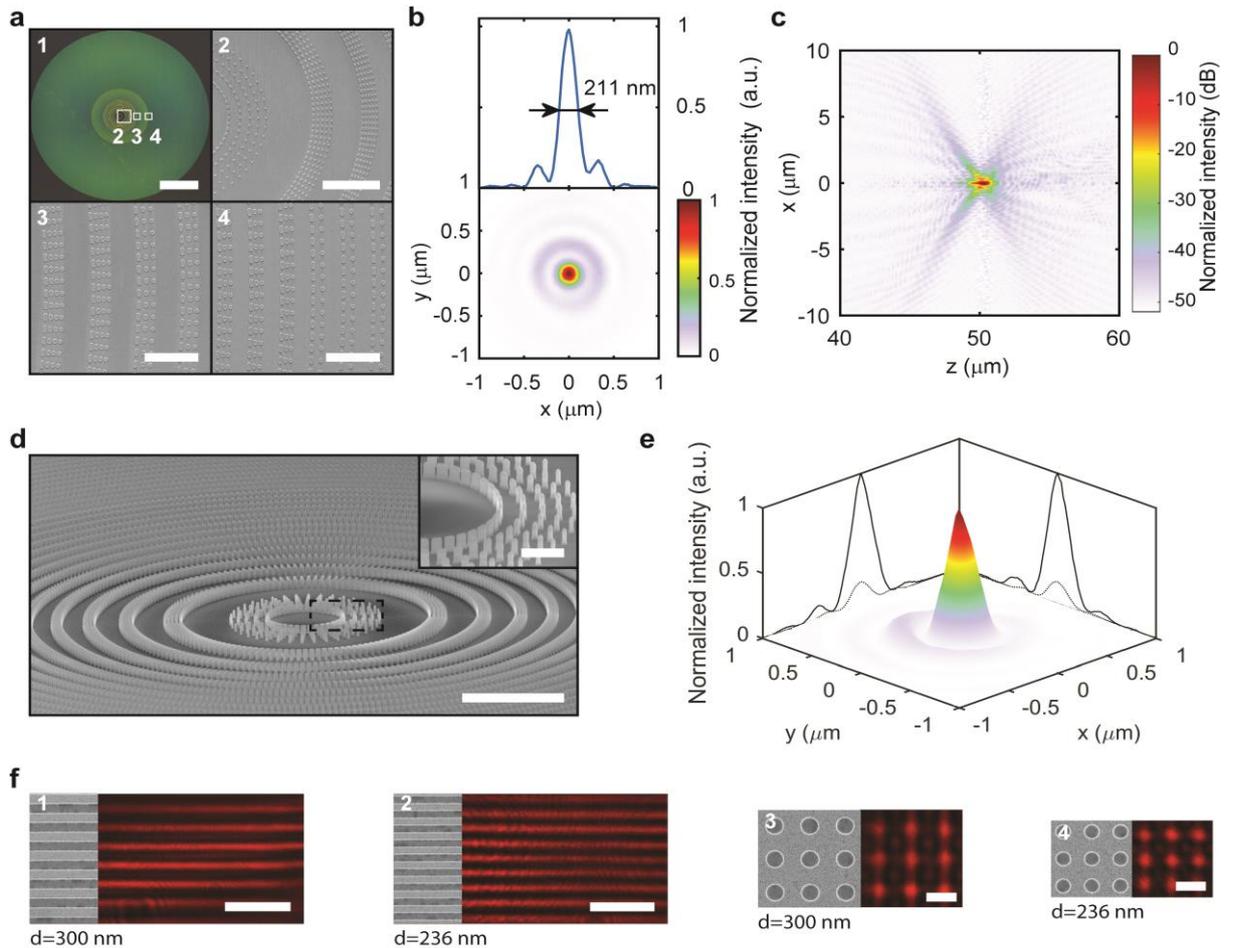

*Figure 4.* High-numerical-aperture immersion metalenses. (a) Bright field image and SEMs of a plasmonic 1.4 numerical aperture (NA) immersion metalens. Scale bars, 50 µm (1), 3 µm (2) and 1 µm (3, 4). The respective locations of the SEMs are indicated by the white squares. (b) Imaged intensity at the focal plane (x-y plane) and the normalized field intensity profile of the plasmonic metalens in (a). (c) Intensity distribution in the meridional plane (x-y plane) of *the plasmonic metalens in dB scale. (d) Tilted SEM of a dielectric amorphous silicon immersion metalens with partial phase control. The NA of the metalens is 1.4 and the focal length is f = 30 µm. Scale bars are, 5 µm and 1 µm for the inset. (e) Measured intensity distribution at the focus of the dielectric metalens at a wavelength of 820 nm. For comparison, the dotted line shows the intensity profile for the plasmonic metalens in (c) normalized to the same input power. (f) Optical images of metallic gratings and nanoholes with gaps of d = 300 nm and d = 236 nm acquired using a plasmonic immersion metalens at a wavelength of 660 nm (the lens has NA = 1.4 and f = 230 µm, see Figure S5, Supporting Information, for experimental set-up). Scale bars, 2 µm (1, 2) and 500 nm (3, 4).*

By arranging the nanoantennas to interfere scattered light constructively at a focal point, similar to previous demonstrations with conventional phase gradient metasurfaces,[6] we created flat lenses. The large angles of deflection attainable with our metasurfaces allow realizing lenses with very high numerical aperture (NA). We fabricated oil immersion metalenses with NA = 1.4 for a design wavelength of 660 nm (see Figure 4a). The measured focal spots are symmetric and show focusing with a full width at half maximum of 211 nm (as shown in Figure 4b and c). With FWHM = 0.32$\lambda$ this is, to the best of our knowledge, the highest resolution metalens built to date. Additionally, a measured Strehl ratio of 0.74 indicates a near diffraction-limited focusing, which is even improved when accounting for the pupil function of the imaging objective (see Figure S3 and Figure S4, Supporting Information, for Strehl ratio calculation and for experimental set-up). In addition to a plasmonic metalens, we used our concept to realize a dielectric high NA immersion lens. This lens consists of closely spaced large aspect ratio amorphous silicon nanopillars as shown in Figure 4d and yields a numerical aperture of 1.4. To be able to couple the scattering at large angles into the substrate we coated the nanostructure with a thin polymer layer with a refractive index close to the one of the glass substrate (See Experimental Section). At an operating wavelength of 820 nm, this dielectric design achieves a symmetric focusing as shown in Figure 4e and a focusing efficiency of 21%, when integrating the focused light intensity in Figure 4e and dividing by the intensity transmitted through an aperture of the same size as the metalens. We tested the performance of a plasmonic 1.4 NA metalens by imaging a resolution test target consisting of metallic gratings and nanoholes. The acquired images in Figure 4f show that our lenses are able to resolve features with distances down to the diffraction limit $d = \lambda/2NA =$ 236 nm at a wavelength of 660 nm.

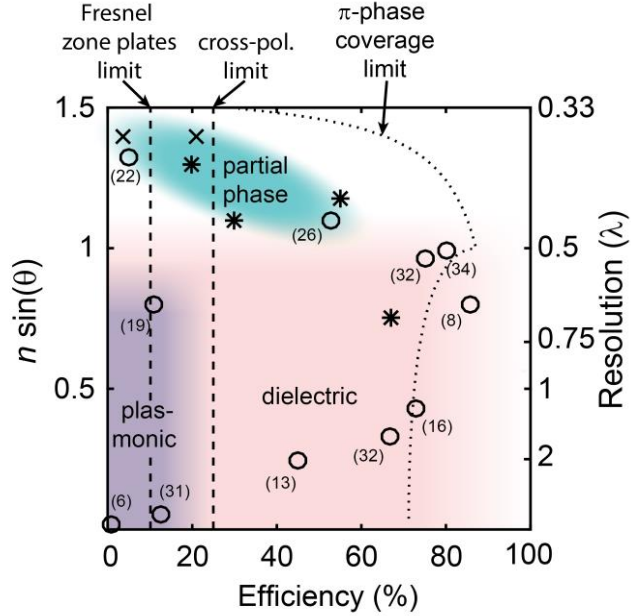

*Figure 5. Anomalous transmission/focusing efficiency vs. deflection angle/numerical aperture. A comparison of different reports of anomalous transmission/focusing efficiency vs. deflection angle/numerical aperture with single layer surfaces in the visible and near infrared wavelength range for PB-phase metasurfaces[6,8,16,19,26], transmittarrays[13,31–34] and Fresnel zone lenses[22]. Numbers in brackets indicate the references. θ and n indicate the maximum achievable angle of anomalous refraction and the refractive index of the medium where the light is anomalously refracted to, respectively (see Figure 1a and c for the schematics). n sin(θ) is a measure of the ability to transfer momentum to light in the direction tangential to the interface, and for metalenses it represents the numerical aperture. The resolution is given as a fraction of the wavelength λ for metalenses assuming the ideal diffraction limited case. Previously reported single-layer plasmonic surfaces have shown efficiencies below the 25% cross-polarization limit (horizontal dashed line),[18] and dielectric surfaces have achieved up to 86% in efficiency.[8] With metalenses the largest numerical aperture demonstrated to date is 1.1.[26] Our experimental results are indicated by star and cross symbols for the beam deflection and lensing cases, respectively, and exceed previously reported results in terms of deflection angle/numerical aperture. The deflection efficiencies of our plasmonic metasurfaces go beyond the previous achievements for plasmonic surfaces, and for dielectric metasurfaces they are comparable to pervious works. The dotted curve shows the theoretical limit of our concept for surfaces with 0-π phase coverage. The vertical dashed lines show the theoretical limit for cross-polarized plasmonic surfaces[18] and for binary Fresnel zone plates.[35]*

**Figure 5** places our results into perspective with respect to the existing state of the art. In comparison to previous reports with full phase control, the results with plasmonic and dielectric metasurfaces based on partial phase control show anomalous refraction of light to significantly larger angles and lensing with higher numerical aperture. In addition, as mentioned earlier (Figure 3a and b), in terms of efficiency, plasmonic surfaces based on our concept substantially outperform previously introduced plasmonic surfaces, while our dielectric surfaces have comparable performance to pervious works with dielectric metasurfaces. Therefore, as evident in Figure 5, this concept markedly advances the performance of metasurfaces to previously inaccessible, new domains. The calculated theoretical limit for surfaces with a phase coverage of 0-π (dotted curve), shows the potential for obtaining performances even beyond what is presented here.

**Conclusion**

In summary, we have demonstrated a new concept for designing optical metasurfaces, which enables efficient broadband anomalous refraction to large angles, surfaces with added functionalities such as wavelength de-multiplexing, and efficient metalensing with very high resolution. Our concept is generally applicable to a broad range of scatterer geometries,[27] materials[28] and illumination schemes.[36] By further optimizing the surfaces, using dielectrics with lower loss[33] and improving the fabrication, we expect to be able to realize high-numerical-aperture optical elements with even higher efficiencies than demonstrated here and for different wavelength ranges. We envision future applications of our design approach in efficient and high-resolution metadevices for microscopy, integrated optoelectronics,[37] sensing, holography[38] and interfacing quantum emitters.[39]

**Methods**
*Optical Characterization*

The fabricated surfaces were characterized with a home-built transmission microscope (schematically shown in Figure S3, Supporting Information) using Fourier plane imaging to quantify the intensity of the transmitted and deflected beams. We use white light from a Xenon lamp, which was filtered using a grating monochromator and then impinged on the sample. The transmitted light was collected using an oil immersion objective (NA = 1.4). For angular distribution studies, the back focal plane (Fourier plane) of the objective was imaged onto the chip of a CCD camera. To determine the absolute transmission efficiency to the anomalously refracted beam, the intensity of the pixels corresponding to the diffraction order was integrated over the $1/e^2$ beam waist and divided by the integrated intensity of the reference beam transmitted through the bare glass surface with an aperture of the same size in place in the image plane. Normalizing the efficiency to the transmission of the bare substrate is often done[13,17,32]. Alternatively, normalizing to free space excludes the intrinsic reflection of the glass substrate, which amounts to about 4%. The beam waist was determined from fitting a Gaussian curve to the projected spot.

*Sample Fabrication*

The surfaces were fabricated on 150 μm thick glass coverslips ($n$ = 1.52). To remove organic residues, the coverslips were cleaned in an ultrasonic bath in acetone, isopropyl alcohol (IPA) and deionized (DI) water, each for 15 min, dried using a $N_2$ gun and subsequently treated with oxygen plasma. A thin layer of silver was evaporated directly onto the glass using an electron beam evaporator. For structures with nanorods of width smaller than 80 nm, a 1 nm thick titanium adhesion layer was necessary to avoid delamination of the structures. However, in other cases, the adhesion layer is avoided in order to reduce adhesion-layer-induced plasmon damping.[40] In a next step, the nanostructure was written in a spin-coated HSQ resist layer using standard electron beam lithography. Before spin coating, the substrate was immersed for 30 s in *Surpass 3000* for surface promotion, rinsed in water for 20 s and dried using a nitrogen gun. The patterned area for each anomalous refraction metasurface is 200×200 μm$^2$. To improve the adhesion of the resist to the bare silver a 5 nm thick layer of $SiO_2$ was sputtered onto the metal coated substrate, which also prevented oxidation of the silver film during the processing. To remove the unexposed resist, the substrate was immersed into a developer solution (AZ351B:$H_2O$, 1:3) for 5 min, then immersed into a diluted solution (AZ351B:$H_2O$, 1:8) for 30 s, rinsed in DI water for 5 min and cleaned with acetone and IPA.

The nanoantenna pattern was transferred to the silver layer using argon ion sputter etching. For the removal of residual resist, the samples were then immersed in buffered hydrofluoric acid (1:7) for 5 s, rinsed in H$_2$O and cleaned in acetone and IPA. To protect the silver nanostructures from oxidizing, a 2 nm thick Al$_2$O$_3$ layer was deposited in a low temperature atomic layer deposition process at 50°C.[41]

Dielectric metasurfaces were fabricated from a 350 nm thick amorphous silicon layer deposited with plasma enhanced chemical vapor deposition. Electron beam lithography was performed on an HSQ resist layer, using a 5 nm sputter deposited SiO$_2$ layer for improving the resist adhesion and using a spin-coated PEDOT:PSS layer for charge dissipation. The resist was developed in an aqueous solution of 1% NaOH and 4% NaCl. The pattern was transferred to the amorphous silicon film using inductively coupled plasma reactive ion etching. To be able to access large scattering angles into the substrate, an 850 nm thick Poly(methyl methacrylate) (PMMA) layer was spin coated on top of the nanostructure.

*Simulations*

The numerical modeling of the nanostructures was carried out using the finite-difference time-domain (FDTD) method. Simulations were performed with a standard commercially available FDTD software (*Lumerical FDTD Solutions*). Tabulated values from Palik[42] were used for the dielectric functions of silver and Al$_2$O$_3$, the optical properties of amorphous silicon were measured with an ellipsometer and a constant refractive index was considered for the PMMA and the glass substrate.

**Acknowledgements**
We would like to thank Dr. Antonis Olziersky and Dr. Yuriy Fedoryshyn for their help with electron beam lithography, Ute Drechsler for advice with the nanofabrication, and Andreas Messner for the ellipsometry measurements. We would also like to express our thanks to Olivier Knutti and Deborah Carneiro Moreira for their help in the start-up phase of the project, and Dr. Thomas Schutzius for fruitful discussions.

# Supporting Information

**Supporting Figures**

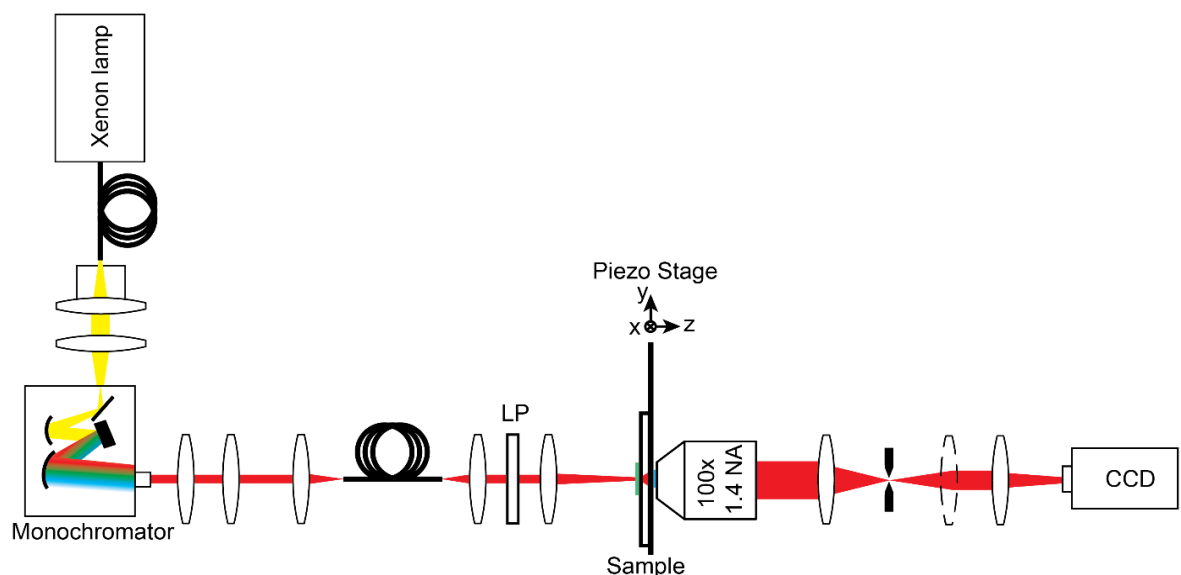

**Figure S1.** Schematic of the optical measurement setup for metasurface characterization. The setup is used either in normal imaging mode (with the dashed line lens in place) or in Fourier imaging mode (by removing the dashed line lens from the optical path). A linear polarizer (LP) is used to set the polarization of the incident light along the nanorod long axis.

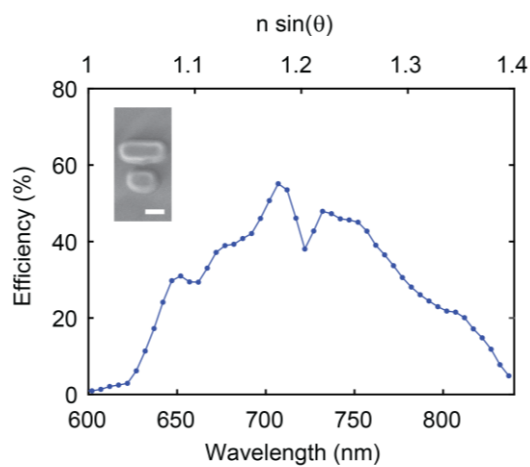

**Figure S2.** (a) Large angle beam deflection with a dielectric metasurface. Measured transmission efficiency of anomalous refraction to large angles by a dielectric two-element array metasurface with partial phase control. The surface consists of an array of 350 nm tall amorphous silicon nanopillars and has a periodicity of $P_x = 600$ nm and $P_y = 300$ nm. The surface is covered with an 850 nm thick layer of PMMA. Deflected light above 810 nm is not entirely captured by the oil immersion objective. The inset shows a scanning electron micrograph of a unit cell of the measured surface. The scale bar indicates a distance of 100 nm.

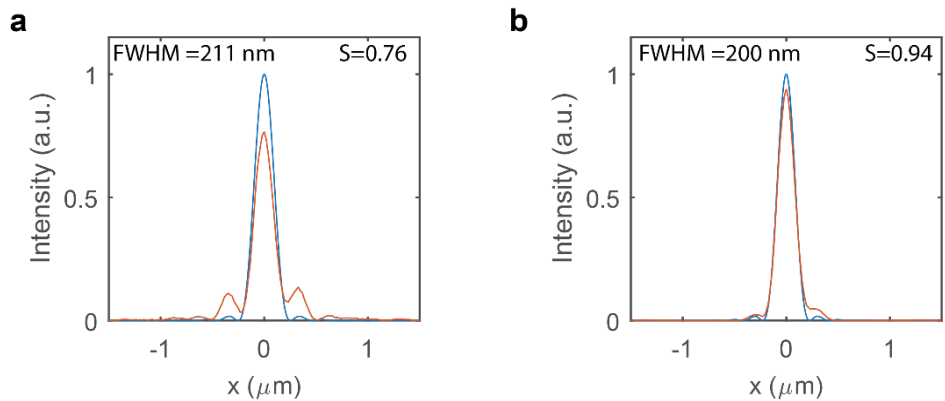

**Figure S3.** Strehl ratio calculation and deconvolution from the pupil function. (a) Measured intensity profile (red curve) and ideal airy disk function (blue curve) at the focal point. A Strehl ratio of $S = 0.76$ is calculated for the 1.4 NA lens indicating a near diffraction limited lensing. The Strehl ratio is determined by the same method as used for dielectric metalenses[8]. (b) The measured intensity profile at the focal spot deconvolved from the pupil function of the objective (red curve) and the ideal airy disk function (blue curve). A corrected Strehl ratio of $S = 0.94$ is calculated after deconvolution. Deconvolution is performed with an ideal airy disk function, as the Strehl ratio of the imaging objective is 0.95.

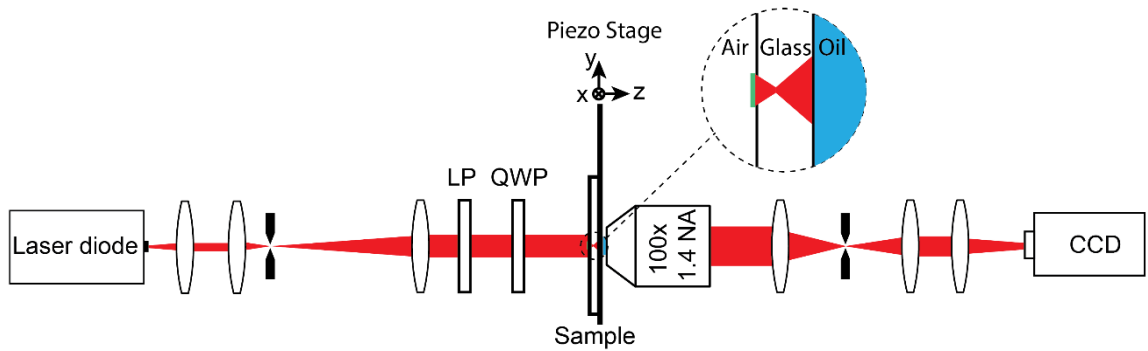

**Figure S4.** Schematic of the optical measurement setup for the immersion metalens characterization. The sample is mounted on a piezo stage, which allows imaging the electric field distribution at different z-positions. A linear polarizer (LP) and quarter wave plate (QWP) are used to obtain a circularly polarized incident light. A laser diode of 660 nm or 820 nm is used for the measurement of the plasmonic lens or dielectric lens, respectively.

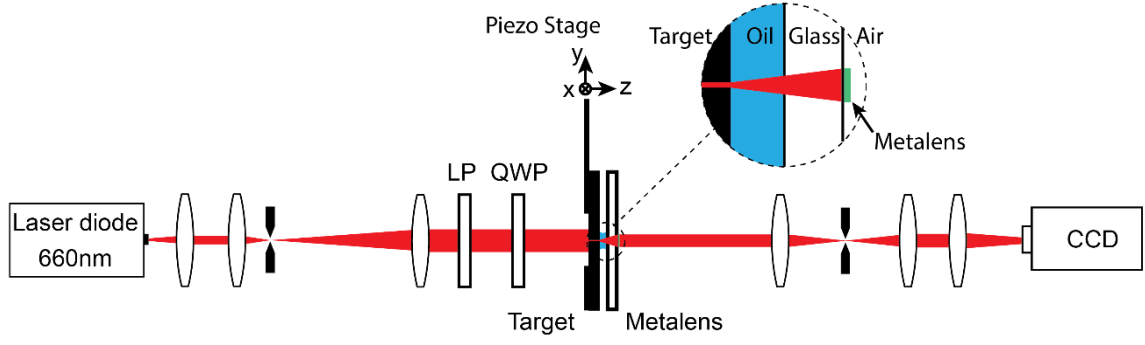

**Figure S5.** Schematic of the optical measurement setup for imaging with an immersion metalens. A resolution target is mounted on a piezo stage and is imaged by the metalens, which is fabricated on a glass piece.

**Supporting Note 1: Partial phase gradient efficiency analysis**

Infinitely thin, ideal phase gradient metasurfaces without polarization conversion can be modeled as a phase grating with the complex electric field amplitude, shown in Figure 1d. Reducing the phase coverage results in the depicted piecewise linear complex electric field amplitude

$$\mathbf{E}(x) = \begin{cases} \mathbf{E}_0 e^{2\pi i \frac{x}{P_x}} ; & 0 \leq x \leq P_x \gamma \\ 0 ; & P_x \gamma \leq x \leq P_x, \end{cases} \quad (2)$$

where $\gamma$ is the phase coverage ratio (fraction of the covered surface area), $x$ is the coordinate along the interface, and $\mathbf{E}_0$ is the electric field amplitude. The grating orders of this phase grating are obtained by

$$E(k_x) = \sum_{m=-\infty}^{\infty} c_m \delta(k_x - m\kappa), \quad (3)$$

where $\kappa = \dfrac{2\pi}{P_x}$ is the spatial frequency of the grating and $k_x$ the x-component of the wavevector. The coefficients $c_m$ are given by

$$c_m = \int_0^{\gamma P_x} \mathbf{E}(x) e^{-ikmx} dx = \frac{\mathbf{E}_0 i}{2\pi + m\kappa P_x} \left( e^{-i(2\pi/P_x + m\kappa)P_x\gamma} - 1 \right). \quad (4)$$

The diffraction efficiency of the different diffraction orders is obtained by

$$\eta_m = \frac{\overline{c}_m \cdot c_m}{\sum \overline{c}_n \cdot c_n}. \quad (5)$$

This results in the diffraction efficiency of the diffraction orders m = -1,0,1

$$\eta_m = \begin{cases} \dfrac{4\pi^2 \gamma^2}{4\pi^2 \gamma^2 + \sin^2(2\pi\gamma) - 2\cos(2\pi\gamma) + 2}; & m = 1, \\ \dfrac{2 - 2\cos(2\pi\gamma)}{4\pi^2 \gamma^2 + \sin^2(2\pi\gamma) - 2\cos(2\pi\gamma) + 2}; & m = 0, \\ \dfrac{\sin(2\pi\gamma)^2}{4\pi^2 \gamma^2 + \sin^2(2\pi\gamma) - 2\cos(2\pi\gamma) + 2}; & m = -1. \end{cases} \quad (6)$$

To account for a non-uniform scattering pattern of the scatterers, the diffraction efficiency can be modified according to

$$\eta_{dm} = I(\theta)\eta_m, \qquad (7)$$

where $I(\theta)$ denotes the scattering pattern as a function of the diffraction angle θ. This scattering pattern is numerically approximated[43], assuming an in-plane electric dipole directly at the interface. Figure S6a shows the resulting diffraction efficiency for a varying phase coverage.

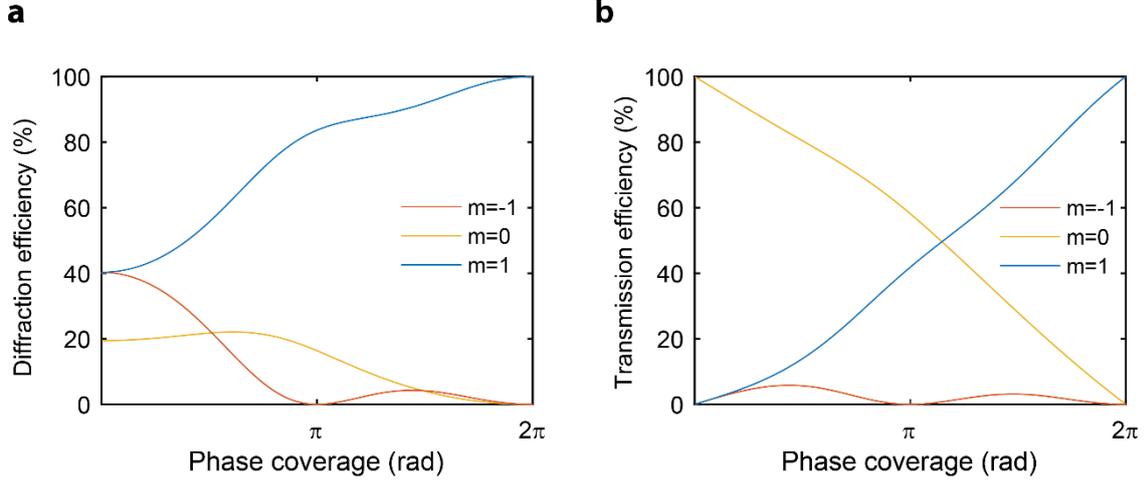

**Figure S6.** Diffraction and transmission efficiency as function of phase coverage. (a) Calculated diffraction efficiency for electric dipoles at an air-glass ($n = 1.5$) interface with a periodicity $P_x = 700$ nm with varying phase coverage. (b) Transmission efficiency of the different orders assuming that the impinging wave interacts with the metasurface according to the fraction of covered surface area γ.

To account for the limited scattering cross section of the metasurface structure (due to limited surface coverage), we approximate the transmission efficiency by assuming that the fraction of light interacting with the metasurface corresponds to the fraction of the covered surface area, i.e. γ. The transmission efficiency of the different orders ($m = -1,0,1$) is then given by

$$\eta_t = \begin{cases} \gamma \dfrac{4\pi^2\gamma^2}{4\pi^2\gamma^2 + sin^2(2\pi\gamma) - 2cos(2\pi\gamma) + 2}; & m = 1, \\ 1-\gamma+\gamma \dfrac{2-2cos(2\pi\gamma)}{4\pi^2\gamma^2 + sin^2(2\pi\gamma) - 2cos(2\pi\gamma) + 2}; & m = 0, \\ \gamma \dfrac{sin(2\pi\gamma)^2}{4\pi^2\gamma^2 + sin^2(2\pi\gamma) - 2cos(2\pi\gamma) + 2}; & m = -1. \end{cases} \qquad (8)$$

The obtained transmission efficiency with varying phase coverage is shown in Figure S6b taking into account also the scattering pattern from Equation (7). As expected, at full phase coverage, the impinging beam is fully transmitted to the anomalous beam ($m = 1$), and for a limited phase coverage of $\gamma = \pi$ less than 50% is transmitted to the anomalous beam, since here we assume that the surface interacts only with half of the impinging wave. This gives a conservative estimate of the maximum attainable transmission efficiency for a given phase coverage. If the scattering cross section is maximized, such that the complete impinging beam interacts with the array, the transmission efficiency will be significantly higher and closer to the diffraction efficiency. An example of such a case is discussed in Figure 3b, where a transmission efficiency of 67% is achieved by only covering the phase range of 0-0.7π using dielectric scatterers.

A comparison to full-wave simulations of a periodic structure of aligned in-plane electric dipoles at an air-glass interface with varying phase coverage shows good agreement with this simplified analytical model (see Figure S7). The maximum transmission efficiency depends largely on the scattering cross section, which is only obtained through more detailed analysis of the scatterers, and requires also including scatterer dependent properties such as backwards scattering and absorption. In addition, even higher efficiencies may be attained by scatterers with both electric and magnetic dipoles.

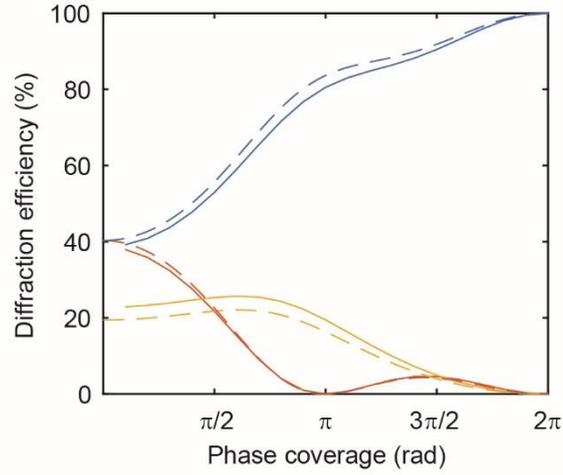

**Figure S7.** Comparison of the analytical model and full-wave simulation. Comparison of the calculated diffraction efficiencies of phase gradient metasurfaces with full-wave simulations (solid line) and based on our analytical model (dashed line). For the simulation ten discrete dipole sources with an imposed linear phase gradient were used at an air-glass ($n = 1.5$) interface with a periodicity $P_x = 700$ nm.